\documentclass[twocolumn,showpacs,amsmath,amssymb,superscriptaddress,floatfix,pra]{revtex4}

\usepackage{amsfonts}
\usepackage{mathrsfs}

\usepackage{amsmath}
\usepackage{graphicx}
\usepackage{bm}
\usepackage{subfigure}

\newcommand{\ket}[1]{\mbox{$|#1\rangle$}}
\newcommand{\bra}[1]{\mbox{$\langle #1|$}}

\begin{document}

\title{Linear optical quantum computation with imperfect entangled photon-pair sources and inefficient non-photon-number-resolving detectors}

\author{Yan-Xiao Gong}
\email{gongyanxiao@gmail.com}
\affiliation{National Laboratory of Solid State Microstructures and Department of Physics, Nanjing University, Nanjing, 210093, People's Republic of China}
\affiliation{Key Laboratory of Quantum Information, University of
Science and Technology of China, CAS, Hefei, 230026, People's
Republic of China}

\author{Xu-Bo Zou}
\email{xbz@ustc.edu.cn}
\affiliation{Key Laboratory of Quantum Information, University of
Science and Technology of China, CAS, Hefei, 230026, People's
Republic of China}

\author{Timothy C. Ralph}
\affiliation{Centre for Quantum Computer Technology and Physics Department, University of Queensland, QLD 4072, Brisbane, Australia}

\author{Shi-Ning Zhu}
\affiliation{National Laboratory of Solid State Microstructures and Department of Physics, Nanjing University, Nanjing, 210093, People's Republic of China}

\author{Guang-Can Guo}
\affiliation{Key Laboratory of Quantum Information, University of
Science and Technology of China, CAS, Hefei, 230026, People's
Republic of China}

\begin{abstract}

We propose a scheme for efficient cluster state quantum computation by using imperfect polarization-entangled photon-pair sources, linear optical elements and inefficient non-photon-number-resolving detectors. The efficiency threshold for loss tolerance in our scheme requires the product of source and detector efficiencies should be $>1/2$ - the best known figure. This figure applies to uncorrelated loss. We further find that the loss threshold is unaffected by correlated loss in the photon pair source. Our approach sheds new light on efficient linear optical quantum computation with imperfect experimental conditions.

\end{abstract}


\maketitle

Linear optical quantum computation (LOQC) has been known to be possible since Knill, Laflamme, and Milburn (KLM) proposed a scheme by employing measurement-induced nonlinearities \cite{KLM2001}. Although scalable in principle, their scheme requires an unacceptably large number of operations. Much progress has been made in recent years to simplify LOQC \cite{review:Kok2007}. One of the approaches is cluster state quantum computation \cite{cluster:firstPRL,cluster:nielsen2004}. This model, different from the circuit computation model, starts from an entangled state, known as a cluster state, and requires only adaptive single-qubit measurements and feed-forward operations. This model has been the subject of numerous LOQC  investigations, such as proposals for scalable cluster state generation \cite{cluster:browne2005,cluster:Duan2005,cluster:bodiya2006,cluster:gilbert2006,cluster:joo2007}, experimental realizations of small-scale cluster quantum computation \cite{cluster:Walther2005,cluster:Prevedel2007,cluster:tame2007,cluster:chen2007,cluster:vallone2008,cluster:Tokunaga2008}, and fault-tolerant thresholds \cite{cluster:thresholds:Nielsen2005,cluster:thresholds:dawson2006PRL,cluster:thresholds:dawson2006PRA,cluster:thresholds:raussendorf2007}.

Error correction is essential for large scale quantum computation. A major error source in LOQC is photon loss, arising from imperfect photon sources, inefficient photon detectors and non-ideal optical circuits. For the circuit computation model, a loss error threshold of between $1.78\%$ and $11.5\%$ was found using the seven-qubit CSS code \cite{loss:Silva2005} and up to $18\%$ with parity encoding \cite{Loss:ralph2005}. Thresholds for loss are considerably higher than for errors such as bit flips because loss in LOQC is a locatable error that removes the qubit, rather than changing its logical value. Even higher loss thresholds have been found for cluster state quantum computation. Recently, by building tree-cluster states, and assuming the loss error is in an independent degraded (ID) form, i.e., each qubit in the cluster state has the same and uncorrelated loss error rate, Vanarva, Browne and Rudolph proposed a scheme \cite{cluster:varnava2008} for loss tolerant LOQC by using imperfect single-photon sources, inefficient photon-number-resolving detectors and linear optical elements, providing the product of the detector efficiency and source efficiency greater than $2/3$. This is the current highest known loss threshold for any LOQC scheme.

In this paper, we consider a different technique for building the tree clusters that removes the requirement for photon-number-resolving detectors and  leads to an improved loss threshold  of $1/2$. Furthermore, the ID source model considered in Ref. \cite{cluster:varnava2008} assumes no correlated loss occurs for the sources of entangled photons. This would seem unlikely to be true as for a generic deterministic model any loss in the pumping process of the entangled state source will lead to a correlated two-photon loss in the output. It is usually assumed that such correlated noise will be detrimental. Here, in contrast, we find that our construction method is fail safe to such correlated noise. We consider a situation in which both correlated and uncorrelated loss errors occur to a polarization-entangled photon-pair source. We find that thresholds are strongly improved compared with that in Ref. \cite{cluster:varnava2008}. In particular, if no uncorrelated loss occurs to either photon, we obtain that the detector efficiency is allowed to be as small as a value larger than $1/2$.

The mixed state from the inefficient source is assumed to be in the form
\begin{align}
  \label{eq:source}\rho_s=&(1-\eta_s-\eta_a-\eta_b)\ket{\text{vac}}\bra{\text{vac}}+\eta_s\ket{\Phi^+}_{a,b}\bra{\Phi^+}_{a,b}\nonumber\\
  &+\frac{\eta_a}{2}\left(\ket{H}_a\bra{H}_a+\ket{V}_a\bra{V}_a\right)\nonumber\\
  &+\frac{\eta_b}{2}\left(\ket{H}_b\bra{H}_b+\ket{V}_b\bra{V}_b\right),
\end{align}
with the target entangled state
\begin{align}
    \label{eq:bell}\ket{\Phi^+}_{a,b}=\frac{1}{\sqrt{2}}(\ket{HH}_{a,b}+\ket{VV}_{a,b}),
\end{align}
where $\eta_s$, $\eta_a$, $\eta_b$ represent the rates of emitting a photon-pair, only one photon in beam $a$, $b$, respectively, with $\ket{H}$ ($\ket{V}$) denoting the horizontal (vertical)
polarization state and $\ket{\text{vac}}$ the vacuum state. Note that, the source model we consider has the property that it emits one and only one photon-pair, one or both of which may be lost with some probability. As in Ref. \cite{cluster:varnava2008}, we do not consider the possibility of higher photon number emission, that means, the source we consider does not emit more than two photons once it emits photons.

Another way to understand the source is in terms of loss error rate. A perfect Bell state given by Eq. (\ref{eq:bell}) is assumed to suffer a correlated loss rate $f_c$, uncorrelated loss rates $f_a$ and $f_b$ in modes $a$ and $b$, respectively. Then the state given by Eq. (\ref{eq:source}) can be equivalently rewritten under the following relations
\begin{equation}
\begin{split}
\eta_s&=(1-f_c)(1-f_a)(1-f_b),\\
\eta_a&=(1-f_c)(1-f_a)f_b,\\
\eta_b&=(1-f_c)(1-f_b)f_a.
\end{split}
\end{equation}

The photon detectors we consider are the realistic detectors commonly used in photonic experiments. The detector can not resolve the number of photons detected, but tell us whether photons exist in a detection event with non-unit probability $\eta_d$. The dark count of this kind of detector is usually very low, and hence here is neglected. The POVM describing a non-photon-number-resolving detector is \cite{PDC:Koka}
\begin{align}
  E_{\text{click}}&=\sum_{n=0}^{\infty}\left[1-(1-\eta_d)^n\right]\ket{n}\bra{n},\\
  E_{\text{off}}&=\sum_{n=0}^{\infty}\left(1-\eta_d\right)^n\ket{n}\bra{n}.
\end{align}

\begin{figure}[tb]
\centering
\includegraphics[width=0.3\textwidth]{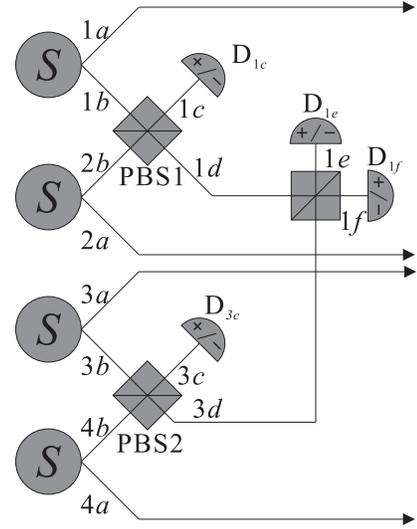}
\caption{Scheme for generating a four-photon GHZ
state from four photon-pair sources (represented by $S$). Lowercase letters
and numbers label the beams. Polarizing beam splitter (PBS)
transmits horizontally polarized ($\ket{H}$) photons and reflects
vertically polarized ($\ket{V}$) photons. $+/-$ denotes $\pm45^{\circ}$ polarization basis.}\label{fig:scheme}
\end{figure}

According to the strategy for creating tree-clusters in Ref. \cite{cluster:varnava2008}, we first need to prepare a four-photon GHZ state. Our scheme for generating a
four-photon polarization-entangled GHZ state using four photon-pair
sources is shown in Fig.~\ref{fig:scheme}, which can be regarded as
two steps. The first step is fusing two sources with a polarizing beam splitter (PBS) followed
by a photon detection at one output port in the basis of $+/-$. Here
$+$ and $-$ denote $45^{\circ}$ and $-45^{\circ}$ polarization
basis, respectively. For example, based on successful detections at
$D_{1c}(D_{1c}^+,D_{1c}^-)$, the unnormalized state in modes $1a$,
$2a$, and $1d$ can be written as
\begin{align}\label{eq:step1}
  &\frac{1}{2}\eta_s^2\eta_d\ket{\text{GHZ}_3}_{1a,2a,1d}\bra{\text{GHZ}_3}_{1a,2a,1d}\nonumber\\
  &+\frac{1}{4}\eta_s\eta_b\eta_d\left(\ket{HH}_{1a,1d}\bra{HH}_{1a,1d}+\ket{VV}_{1a,1d}\bra{VV}_{1a,1d}\right.\nonumber\\
  &\left.+\ket{HH}_{2a,1d}\bra{HH}_{2a,1d}+\ket{VV}_{2a,1d}\bra{VV}_{2a,1d}\right)\nonumber\\
  &+\frac{1}{4}\eta_b^2\eta_d\left(\ket{H}_{1d}\bra{H}_{1d}+\ket{V}_{1d}\bra{V}_{1d}\right)+\cdot\cdot\cdot,
\end{align}
where
\begin{align}
    \ket{\text{GHZ}_3}_{1a,2a,1d}=\frac{1}{\sqrt{2}}\left(\ket{HHH}_{1a,2a,1d}+\ket{VVV}_{1a,2a,1d}\right).
\end{align}
Here, for simplicity, we neglect the amplitudes containing no photon in beam $1d$. Note that, if the detector $D_{1c}^-$ clicks a correction of $\sigma_z$ operation on beam $1d$ ($1a$, or $2a$) is needed (analogous calculations can be found in Refs. \cite{CNOT:Pittman2001,gong:parity}). It is clear that the state in modes $3a$, $4a$, and $3d$ has the same form as
Eq.~(\ref{eq:step1}).

The second step is fusing the two mixed states obtained in the first step with a PBS followed by a photon detection
in each output port. Based on twofold coincidence detection at
detectors $D1e(D1e^+,D1e^-)$ and $D1f(D1f^+,D1f^-)$, the normalized state in modes $1a$, $2a$, $3a$, and $4a$ is in the following form
\begin{align}\label{eq:step2}
  \rho_r=&(1-\epsilon)^4\rho_0+(1-\epsilon)^3\epsilon\rho_1+(1-\epsilon)^2\epsilon^2\rho_2\nonumber\\
  &+(1-\epsilon)\epsilon^3\rho_3+\epsilon^4\ket{\text{vac}}\bra{\text{vac}},
\end{align}
where $\rho_0$ is a perfect four-photon GHZ state,
\begin{align}\label{eq:ghz4}
  &\ket{\text{GHZ}_4}_{1a,2a,3a,4a}\nonumber\\&=\frac{1}{\sqrt{2}}\left(\ket{HHHH}_{1a,2a,3a,4a}+\ket{VVVV}_{1a,2a,3a,4a}\right),
\end{align}
 with $\rho_1$, $\rho_2$, and $\rho_3$ representing the mixed states evolved from a GHZ state by losing one, two, and three photons, respectively. Explicitly,
\begin{align}\label{eq:s123}
\rho_1&=\sigma_{1a,2a,3a}+\sigma_{1a,2a,4a}+\sigma_{1a,3a,4a}+\sigma_{2a,3a,4a},\nonumber\\
\rho_2&=\varrho_{1a,2a}+\varrho_{1a,3a}+\varrho_{1a,4a}+\varrho_{2a,3a}+\varrho_{2a,4a}+\varrho_{3a,4a},\nonumber\\
\rho_3&=\varsigma_{1a}+\varsigma_{2a}+\varsigma_{3a}+\varsigma_{4a}
\end{align}
where,
\begin{equation}
\begin{split}
\sigma_{i,j,k}&=\frac{1}{2}\left(\ket{HHH}_{i,j,k}\bra{HHH}_{i,j,k}\right.\\
&\ \ \ \ \ \ \ \ \ \ \left.+\ket{VVV}_{i,j,k}\bra{VVV}_{i,j,k}\right),\\
\varrho_{i,j}&=\frac{1}{2}\left(\ket{HH}_{i,j}\bra{HH}_{i,j}+\ket{VV}_{i,j}\bra{VV}_{i,j}\right),\\
\varsigma_{i}&=\frac{1}{2}\left(\ket{H}_{i}\bra{H}_{i}+\ket{V}_{i}\bra{V}_{i}\right),
\end{split}
\end{equation}
with $i,j,k=1a,2a,3a,4a$.

From the above four equations, we can see the resource state produced, $\rho_r$, has an ID form as in Ref. \cite{cluster:varnava2008}, with the ID loss probability
\begin{align}\label{eq:idloss}
  \epsilon=\frac{\eta_b}{\eta_b+\eta_s}=f_a.
\end{align}
The success probability to obtain $\rho_r$ is
\begin{equation}\label{eq:success}
  P_{\text{success}}=\frac{1}{8}\eta_d^4\left(\eta_b+\eta_s\right)^4=\frac{1}{8}\eta_d^4\left(1-f_c\right)^4\left(1-f_b\right)^4.
\end{equation}
Note that, here when calculating the success probability we have considered the feedforward corrections depending on different detection results (for details on analogous calculations see Refs. \cite{CNOT:Pittman2001,gong:parity}).

\begin{figure}[tb]
\centering
\includegraphics[width=0.45\textwidth]{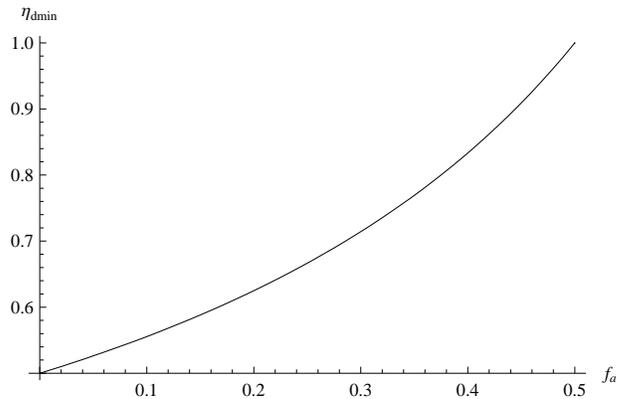}
\caption{Threshold of the detection efficiency ($\eta_{d\text{min}}$) against the uncorrelated loss rate $f_a$ in mode $a$. The detection efficiency $\eta_d$ requires larger than $\eta_{d\text{min}}$. }\label{fig:threshold}
\end{figure}

Next, with $\rho_r$ as a resource, using the method introduced by Varnava-Browne-Rudolph \cite{cluster:varnava2008} (see Fig. 1 in Ref. \cite{cluster:varnava2008}), arbitrarily large tree-cluster states can be built with local Hadamard gates and type-II fusion gates. Therefore, based on Varnava-Browne-Rudolph's error threshold \cite{cluster:loss:varnava2006}, to realize loss tolerant LOQC we require \mbox{$(1-\epsilon)\eta_d>1/2$}, which implies \mbox{$\eta_s\eta_d/(\eta_b+\eta_s)>1/2$}. If we define a ratio \mbox{$\kappa\equiv\eta_b/\eta_s$}, then it is clear that the requirement becomes \mbox{$\eta_d>(\kappa+1)/2$}, which makes sense when \mbox{$\kappa<1$}, i.e., $\eta_b<\eta_s$. A consequent result is that the threshold for the detector improves as $\kappa$ decreases and a max value is $1/2$ when $\kappa$ equals $0$. In the perspective of loss error rate, the requirement becomes
\begin{equation}
(1-f_a)\eta_d>\frac{1}{2},
\end{equation}
which makes sense as long as \mbox{$f_a<1/2$}. Note that not only is the requirement better than the previous threshold of $2/3$, but that it does not depend at all on any correlated loss or loss in arm $b$ of the source. Moreover, as shown in Fig. \ref{fig:threshold}, the detector efficiency requirement gets looser as $f_a$ become smaller. If \mbox{$f_a=0$}, i.e., photon $a$ does not suffer any uncorrelated loss, we get the loosest requirement, \mbox{$\eta_d>1/2$}. Note that $f_c$ and $f_b$ are only required smaller than $1$ to make sure $P_{\text{success}}>0$.

To further compare our threshold with that obtained in Ref. \cite{cluster:varnava2008}, recall that their threshold is \mbox{$\eta_s\eta_d>2/3$}, implying that \mbox{$\eta_d>2/3$} is necessary. While in our scheme, in the case of \mbox{$\kappa<1/3$} or equivalently \mbox{$f_a<1/4$}, the threshold for the detector is certainly improved compared with their threshold. Moreover, photon-number-resolving detectors required in their scheme are not necessary in our scheme.

Consider the case of \mbox{$\eta_b=0$}, or equivalently, \mbox{$f_a=0$}, in which we have the best threshold as just discussed. From Eq. (\ref{eq:step2}) we can see that in this case we can produce a pure four-photon polarization-entangled GHZ state heralded by a fourfold coincidence detection at detectors $D1c$, $D3c$, $D1e$ and $D1f$. As the type-II fusion gate is inherently loss tolerant and does not require photon-number-resolving detectors \cite{cluster:browne2005,cluster:varnava2008}, we can create an arbitrarily large heralded pure tree-cluster state. Hence, our scheme also provides a way to produce ``event-ready'' multiphoton polarization-entangled state \cite{herald:93,herald:07,heralded:niu}.

Eq. (\ref{eq:idloss}) shows that the ID loss rate is exactly the uncorrelated loss occurring to the photon in mode $a$. This implies that this uncorrelated loss $f_a$ can be shifted to the detector. In this respect, we get an equivalent circuit in which the sources have no uncorrelated loss in either photon and the detectors have an effective efficiency $\eta'_d$ satisfying \mbox{$\eta'_d=(1-f_a)\eta_d$}. Consequently, under detection with these ``new'' detectors, pure GHZ states and tree-cluster states can be produced and the threshold is obtained as \mbox{$\eta'_d>1/2$}. Even in the absence of correlated loss this is a superior threshold to that of Vanarva et al \cite{cluster:varnava2008}.

A surprising result is that only the ratio of the generation probability of photon $a$ to that of the photon-pair contributes to the threshold requirement. From the loss error perspective, only the loss rate of either photon affects the threshold. This result may facilitate the experimental realization, as the loss rates on each arm are usually not equal, and thus, to achieve a better threshold for the detector, we can select the modes to satisfy $\eta_b\leq\eta_a$, or equivalently, $f_a\leq f_b$.

Finally, we would like to make some comments on the photon-pair sources we employ. Our source is a bit stricter than the single-photon source usually assumed in LOQC and has not been realized in experiment yet. However, entangled photon pairs are not only a vital resource in quantum optics \cite{book:milburn}, but also of importance in quantum information processing \cite{book:Bouwmeester2000}. Much progress has been made in generating polarization-entangled photon-pairs with the required characteristics, for example, in quantum dot systems \cite{source:entangled:benson2000,source:entangled:stevenson2006,source:entangled:akopian2006,source:entangled:hafenbrak2007,source:entangled:johne2008,source:entangled:young2009}.  We believe that the proof-in-principle experimental demonstration of our scheme is possible with current experimental conditions. In addition, from our analysis above, we can see that, the photon-pair sources in some sense relax the requirement of the photon detectors on number resolving ability and efficiency. From this point of view, our assumption on the source should make sense, as at present photon-number-resolving detectors are more challenging than non-photon-number-resoling detectors. We guess that the source may also find applications in some other LOQC model, for example, in parity-encoded quantum computation \cite{parity:gilchrist2007,parity:hayes2008}.

In summary, we have presented a scheme for loss tolerant LOQC by using imperfect entangled photon-pair sources, inefficient non-photon-number-resolving detectors and linear optical elements. We obtained a better efficiency threshold for loss tolerance of $1/2$ compared to the current best threshold of $2/3$. We have also discussed the roles of correlated and uncorrelated loss errors in the threshold for the photon detector. Our approach opens a new door to efficient LOQC with imperfect experimental conditions.

\emph{Note added.} ---During preparing this paper, we became aware of a related paper \cite{cluster:wei2009}, in which the authors proposed to prepare a three-photon GHZ state using a similar approach to ours. Since building tree-cluster states requires four-photon GHZ states as resource states, they have to fuse two three-photon GHZ states to obtain a four-photon GHZ state with type-II fusion gates, and therefore, our scheme should be more economical and more efficient.

\begin{acknowledgments}
Y.X.G. thanks Gong-Wei Lin for helpful discussions. This work was funded by National Fundamental Research Program
(Grant No.2009CB929601), National Natural
Science Foundation of China (Grants No. 10674128, No. 60121503 and No. 10534020), Innovation Funds from Chinese Academy of Sciences, ``Hundreds of Talents'' program of Chinese Academy of Sciences, Program for New Century Excellent Talents in University, and by the National Key Projects for Basic Researches of China (No. 2006CB921804). T.C.R. is supported by the Australian Research Council.

\end{acknowledgments}

\bibliography{E:/study/latex/bib/ref}
\end{document}